\documentclass[a4paper]{jpconf}
\usepackage{graphicx}
\begin{document}
\title{Gamma-ray variability and correlation properties of blazars observed
with Fermi LAT}

\author{S. Larsson on behalf of the Fermi LAT collaboration and many 
multiwavelength collaborators}

\address{Department of Astronomy, Stockholm University and The Oskar Klein Centre
for Cosmoparticle Physics, Stockholm, Sweden}

\ead{stefan@astro.su.se}

\begin{abstract}
The Fermi Large Area Telescope (Fermi LAT) provides long term systematic monitoring observations of the gamma-ray emission from blazars. The variability properties and the correlation with other wavelength bands are important clues for the evaluation of blazar models. We present results from timing and multiwavelength correlation analysis and discuss differences between blazar classes. 
\end{abstract}

\section{Introduction}
More than
60 AGNs, almost all blazars, were detected by EGRET on Compton GRO, which 
established these sources as a powerful class of gamma-ray emitters
[1]. In its first year of operation Fermi LAT has 
increased the number of known 
gamma-ray blazars by a factor of 10. Even more importantly the instrument
is mapping the full sky every three hours, which allow regular monitoring 
of these sources on time scales from hours to
years. These monitoring observations now form an important part of
many ongoing efforts to study the variability and multiwavelength
properties of blazars. Such observations contains 
information about the relative origin of different spectral components
and about dynamical and radiation processes in blazar jets.

\section{The Fermi Large Area Telescope}

Fermi (the Fermi Gamma-ray Space Telescope) was launched in June 2008 from Cape Canaveral in Florida. The main instrument onboard Fermi is the LAT (Large Area Telescope) which is sensitive to gamma-rays in the energy range 30 MeV to 300 GeV [2]. With a much larger field of view and with a much improved energy resolution compared to its predecessors, Fermi is now providing unprecedented data for studies of active galactic nuclei and other gamma-ray sources. Fermi is operated primarily in a sky survey mode where the full sky is mapped  every three hours. Since LAT has a very wide field of view (about 20\% of the sky) the exposure on any particular sky position is large. Pointed observations are executed only to follow up GRBs or during other exceptional outbursts. The sky surveying and wide field of view of the instrument also mean that the potential for new, serendipitous discoveries is very large. The angular resolution, which increases with photon energy, is about 1 degree at 1 GeV, while source localizations are typically 0.1 degrees.

\section{LBAS, 1LAC and 2LAC}

Based on the first 3 months of observations a LAT bright-AGN Source list (LBAS) 
was produced by the Fermi LAT collaboration [3]. This list contains 106 high confidence
associations (58 FSRQs, 42 BL Lacs, 2 radio galaxies and 4 of unknown type). These were sources with detection test statistic, TS $>$ 100, corresponding 
approximately 
to $10 \sigma $ detections.
By comparison, the First LAT AGN catalog, 1LAC [4] based 
on 11 months of data included already 663 high confidence associations 
(sources with TS $>$ 25, or approximately $5 \sigma $ ). In the recently
released second AGN catalog, 2LAC, this number of sources was increased 
by 50$\% $[5]. 

\section{Blazar variability}

Blazar light curves are often dominated by strong flares. 
The origin of these flares can e.g. be the bulk injection of 
new particles into the jet or strong internal shocks. If the 
variability is made up of shot pulses of different lengths 
and amplitudes produced by such events, the Power Density 
Spectrum (PDS) is determined by the shapes and amplitudes 
of the pulses and by correlations of their relative 
distribution in time. While pulse shapes might be associated 
with, e.g., cooling or light travel time scales, pulse correlations 
will contain information on the processes that are responsible 
for creating the pulses, which could be episodes of strong activity 
near the central black hole. By calculating the PDS, the
Structure Function (SF), and other statistical 
mesures, we can test alternative mechanisms to produce the 
observed variability, keeping in mind that the same PDS or 
SF can be produced by more than one stochastic process.

The Fermi LAT data can be used to study variability
on time scales of hours to years, to measure spectral variations 
and to catch flaring sources with high efficiency.
An example of the capability to study flux and spectral variations
on subday time scales is given by the analysis of the recent
outburst of 3C454.3 [6].

For timing analysis the regular sampling provided by Fermi LAT is 
a great advantage. The initial aim of our study of Fermi blazars
is a characterization the gamma-ray variability. The tools that have 
been used for this are,
\begin{itemize}
\item $\chi ^2$ and excess variance to quantify the presence of variability and its significance.
\item Duty cycle analysis
\item Auto correlation function (ACF)
\item Structure function (SF)
\item Power Density Spectra (PDS)
\item Flare shape analysis
\end{itemize}

Even though the ACF, SF and PDS contain the same information, this 
information is expressed differently, which makes it 
useful to compute all three.
The direct fitting of flare shapes is based on the assumption
that individual flare pulses can be identified. The
advantage with this approach is the possibility to extract
light curve phase information, which is lost in the calculations of 
ACF/SF/PDS. This includes e.g. time asymmetry.

First results of our work, using 11 months of data and
light curves with 3 to 7 days binning for the
LBAS sources was recently published [7] and is partly 
summarized here. 

\subsection{Power Density Spectra}

Power Density Spectra (PDS) of blazars have previously
been studied primarily for optical and radio data and have then 
in general been described by a power law 
$P(\nu) \propto \nu^{-\alpha}$ with $\alpha$ typically in the range
1 to 2. 

For the PDS analysis of Fermi blazars, light curves with 3 day 
binning was used for the 15 brightest LBAS sources
(9 FSRQs and 6 BL Lacs). For a further 13 FSRQs with 
slightly lower flux, the binning was increased to
4 days. A standard Fourier transform was used to compute
power density spectra (PDS) and the white noise level
corresponding to the estimated measurement errors was subtracted.

The power spectral density is normalized to fractional variance
per frequency unit
(\mbox{\it\,rms$^{2}$~I$^{-2}$~Day$^{-1}$}, where $~I$ is the mean flux) 
and the PDS points are
averaged in logarithmic frequency bins. An example PDS for
3C273 is shown in Figure 1 and the 1-week-binned light curve 
for the same source is presented in Figure 2.

\begin{figure}[h]
\begin{minipage}{17pc}
\includegraphics[width=16pc]{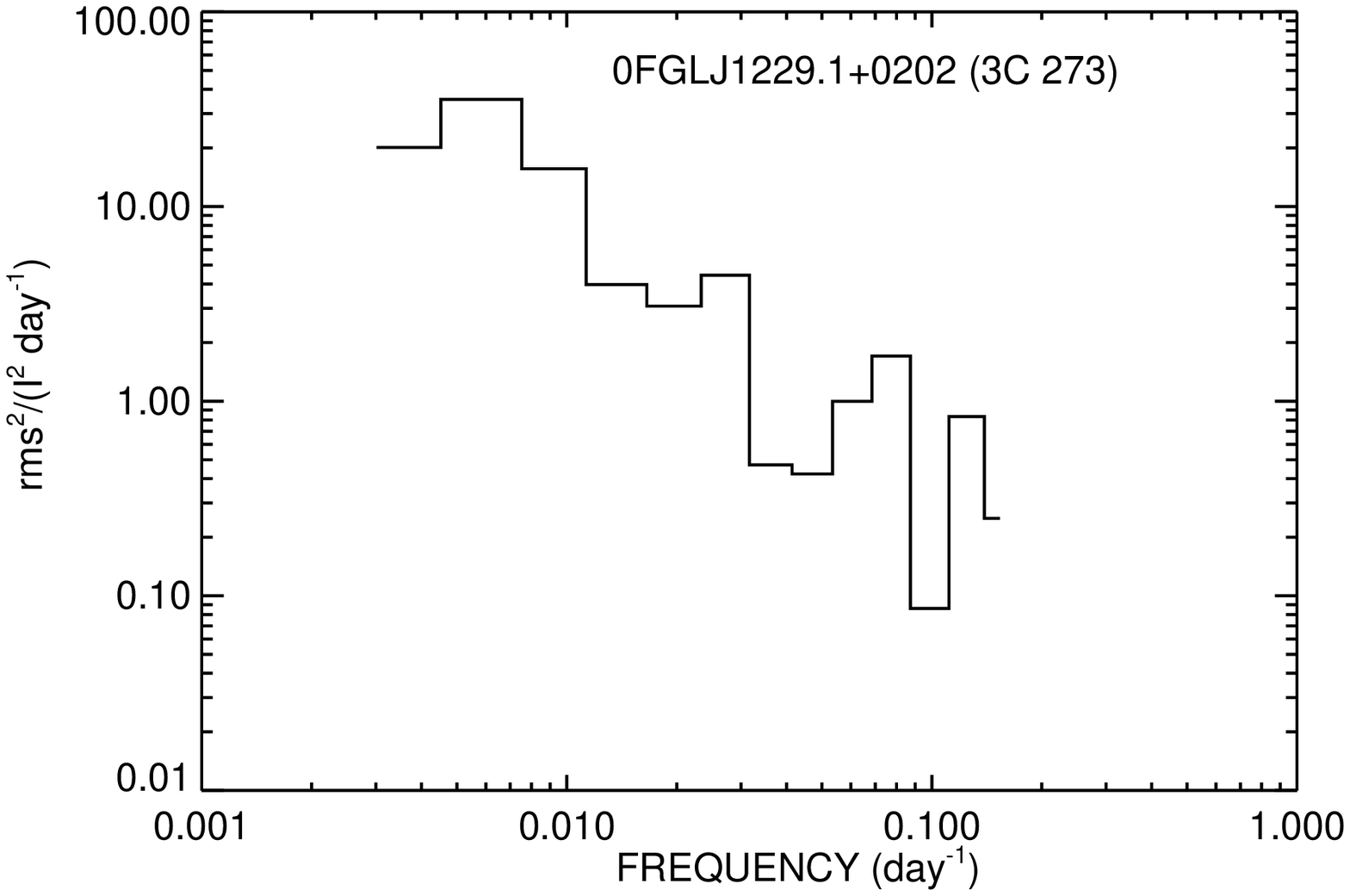}
\caption{\label{label}PDS of the 11 month Fermi light curve for 3C273.
A 3 day binned light curve was used. The PDS is normalized
to variance divided with the square of the mean flux and per frequency
unit (from [7]).}
\end{minipage}\hspace{2pc}%
\begin{minipage}{18pc}
\includegraphics[width=22pc]{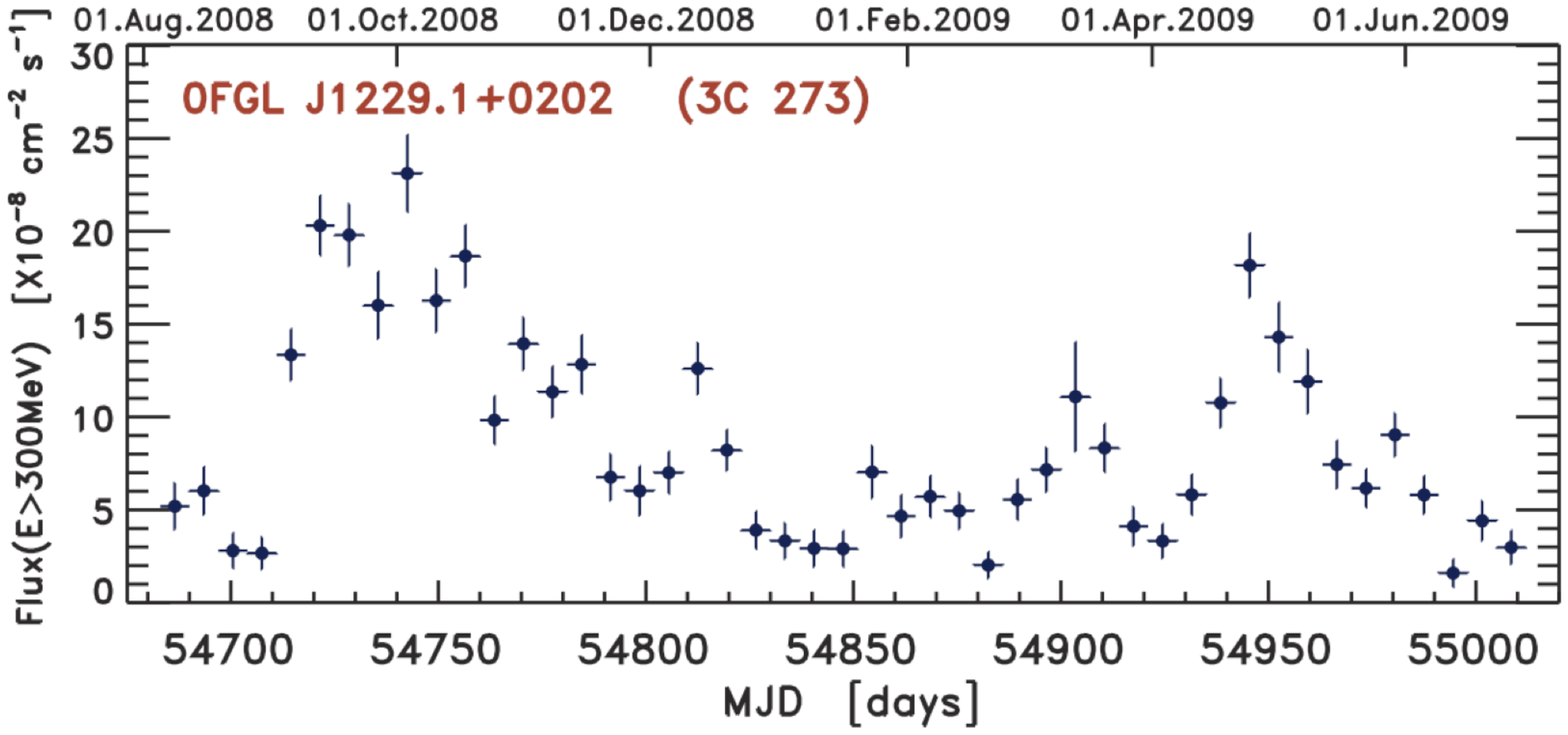}
\caption{\label{label} Fermi 11 month light curve (1 week binning)
for 3C273 (from [7]).}
\end{minipage} 
\end{figure}

There are a number of effects that can distort the shape of the 
computed PDS and cause it to differ from the 'true' intrinsic 
PDS describing the variability of the source. These effects are,

\begin{itemize}
\item Stochastic variability in a time
limited observation.
\item Effects of the observational data and the analysis (e.g. aliasing).
\item Systematic errors
\item Statistical (measurement) errors.
\end{itemize}

For the analysis presented here the last effect dominates at high
frequencies and the first effect at the lower frequencies
where source variance is much larger than the measurement noise.
The white noise level used in the analysis has a significant
effect on the slope deduced from the PDS fit. This 
means that the result is dependent of a reliable estimate of
the measurement errors. These errors were therefore also checked by
comparing some light curves with the corresponding ones obtained by direct
aperture photometry, for which Poisson statistics is valid. This showed
that the uncertainty
in error estimates is not a significant problem for the brightest
sources. For the less bright ones, including all the BL Lacs, this
effect may however, introduce a systematic bias in the PDS slope. 
The effect
was estimated by repeating the analysis for a range of possible
white noise levels and also by analysis of light curves extracted
with different time bins (from 1 to 7 days).
To limit the effect of the white noise on the determination
of the PDS slope, the fits only used frequencies up to 
0.02 \mbox{\it\,day$^{-1}$}.

Observational and instrument systematics were investigated by
analyzing pulsar light curves extracted from the 11 month data with the same
procedure as for the blazars. The most prominent effect is a
periodic modulation that is identified with the 54 day precession period
of the Fermi satellite orbit. In the PDS for individual blazars
this peak is often hidden by the stochastic variability but does show up
when averaging the PDS of a number of sources. The frequency bin at
this period was not used when PDS slopes were estimated.

With the available data the uncertainty in PDS shape
for individual sources is still dominated by stochastic variability.
To reduce this and the statistical fluctuations we can 
average the PDS for a group of sources  under the assumption that the
differences in PDS shape is small compared to the random
fluctuations expected due to the action of the (presumed) underlying
stochastic process. For the 9 brightest FSRQs this averaged
PDS is shown in Figure 3. For frequencies below 0.017 we obtain 
a best fit slope of 1.4 $\pm$0.1. The same PDS is also shown
in Figure 4 together with the corresponding averaged PDS for
the 9 BL Lacs and the 13 moderately bright FSRQs. The differences
in power-law slope are not statistically significant (errors are
estimated from the scatter within each group of sources).

\begin{figure}[h]
\begin{minipage}{17pc}
\includegraphics[width=17pc]{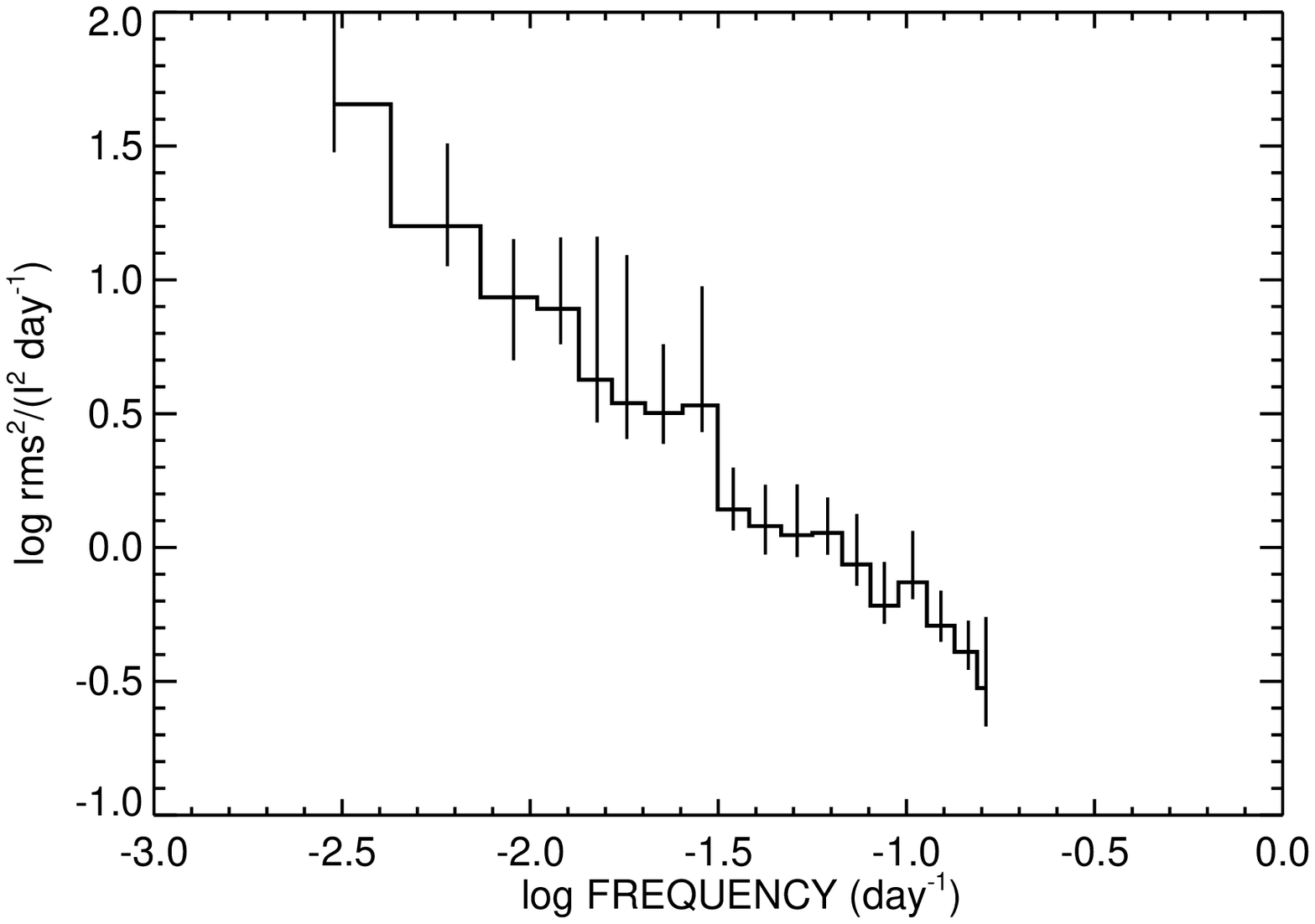}
\caption{\label{label}Average Power Density spectrum, PDS, for the 9 brightest
FSRQs. White noise level based on light curve error estimates
has been subtracted. The error bars are asymmetric 1 sigma errors
of the mean. Our best fit estimate is a PDS slope of
1.4 $\pm$0.1 (from [7]).
\vspace{0.5cm}}
\end{minipage}\hspace{3pc}%
\begin{minipage}{17pc}
\includegraphics[width=17pc]{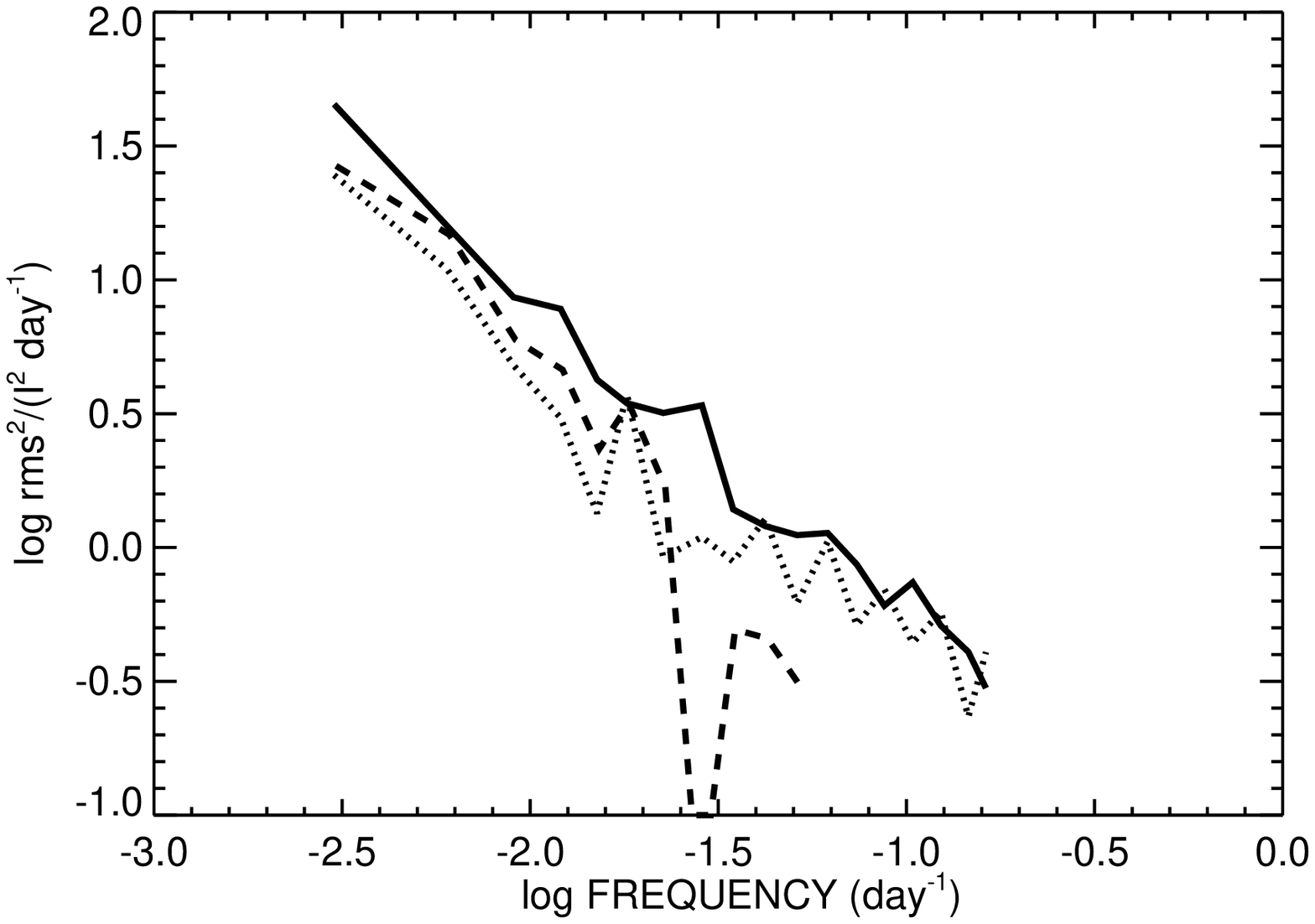}
\caption{\label{label}Comparison of the averaged PDS for three sets of sources,
the 9 bright FSRQs from the upper plot (solid line), the 6 brightest BL Lac's
(dotted line) and 13 additional FSRQs with TS $>$ 1000 (dashed line).
Best fit slope for the BL Lac and fainter FSRQs is  1.7 $\pm$0.3
and  1.5 $\pm$0.2 respectively (from [7]).}
\end{minipage} 
\end{figure}

Since the number of sources is small and the length of observations
relatively short it is too early to draw firm conclusions 
from these results.
It is nonetheless interesting to note a difference between the low
and the high synchrotron peaked BL Lacs (LSP and HSP resp). While the average
slope for the BL Lacs is at least as steep as that of the FSRQs,
the two HSPs, Mkr 421 and PKS2155-305, both have PDS slopes 
flatter than 1.0 (Figure 5). Mkr 421 is the brightest soft X-ray
blazar and a PDS for long time scales at those energies can be
computed from the ASM light curve. The PDS is well fitted by
a power law with index close to -1, as shown in Figure 6.

\begin{figure}[h]
\begin{minipage}{17pc}
\includegraphics[width=17pc]{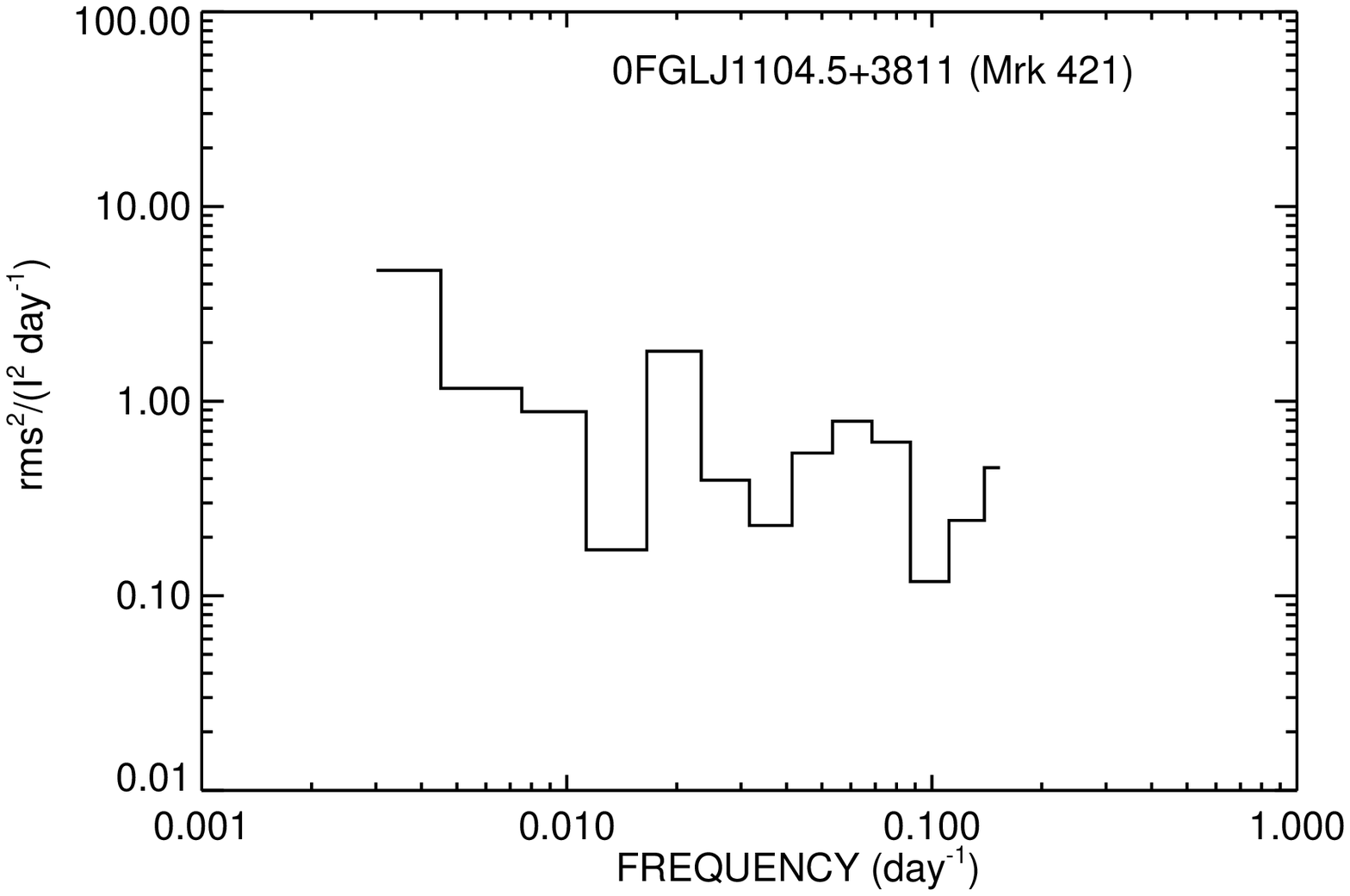}
\caption{\label{label}Gamma-ray PDS for the 11 month Fermi
light curve of the HSP BL~Lac Mkr 421 (from [7]).}
\end{minipage}\hspace{3pc}%
\begin{minipage}{17pc}
\includegraphics[width=17pc]{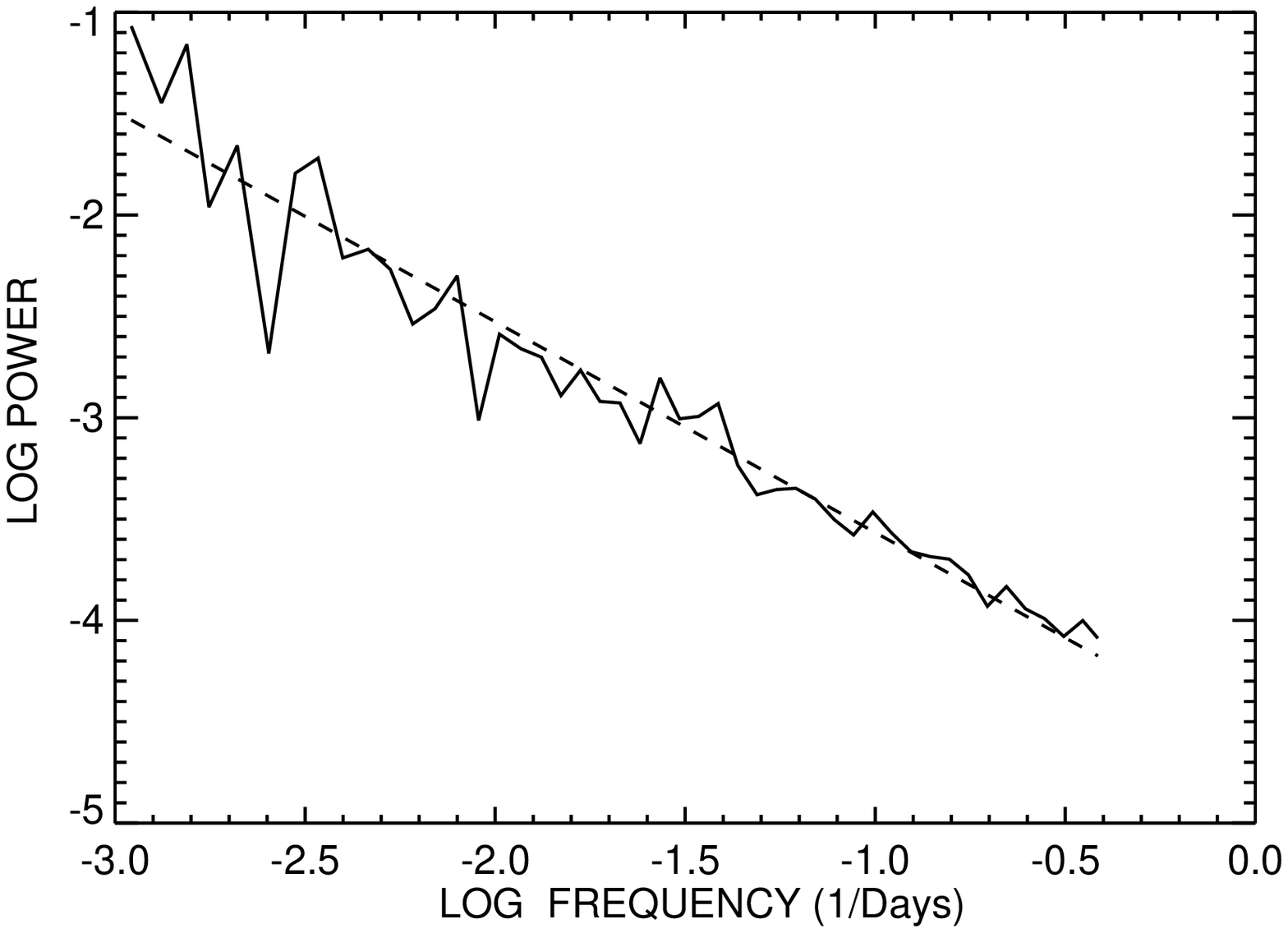}
\caption{\label{label}PDS of the soft X-ray ASM light curve for
Mkr~421 (12 years of data). The PDS power law slope = -1.04 $\pm$0.04. }
\end{minipage} 
\end{figure}

When sources are in a high state it is possible to use
light curves with smaller time bins and to compute a significant
PDS to higher frequencies. In Figure 7 we have combined a
PDS for the time around the December 2009 outburst of
3C454.3 with its 11 month LBAS PDS [8].
The fact that the two PDS join each other
smoothly at the overlapping frequencies implies that the
fractional variability during the bright outburst is similar 
to that of the longer time scale. The same conclusion was drawn
when an intense flaring episode for 3C273 was compared to
its 11 month LBAS PDS [9].  

\begin{figure}[h]
\begin{minipage}{17pc}
\includegraphics[width=17pc]{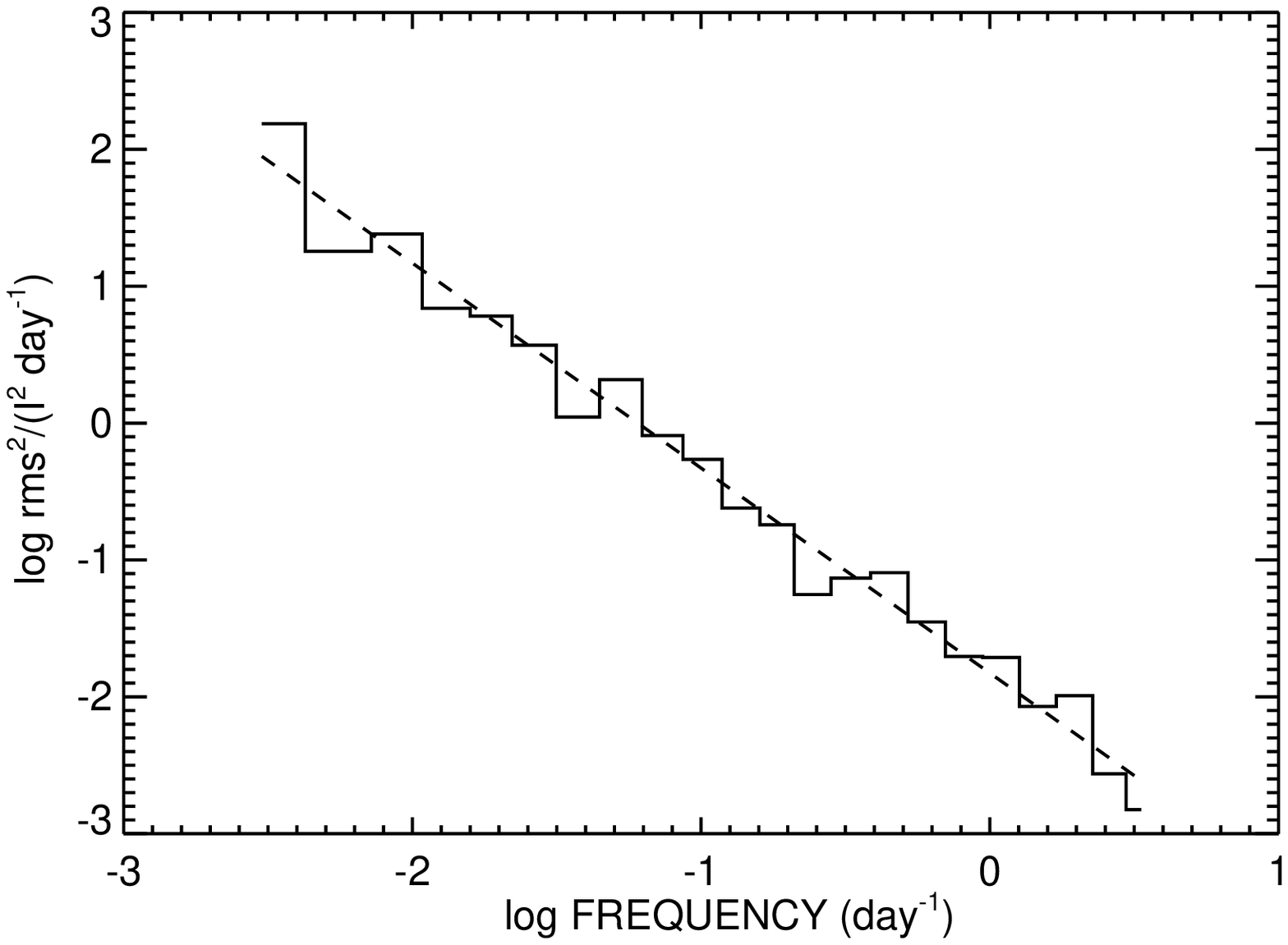}
\caption{\label{label}PDS for the FSRQ 3C454.3 combining
the PDS for the 11 month LBAS data and that of the December 2009
outburst (from [8]).}
\end{minipage}\hspace{3pc}%
\begin{minipage}{17pc}
\vspace{3pc}
\includegraphics[width=17pc]{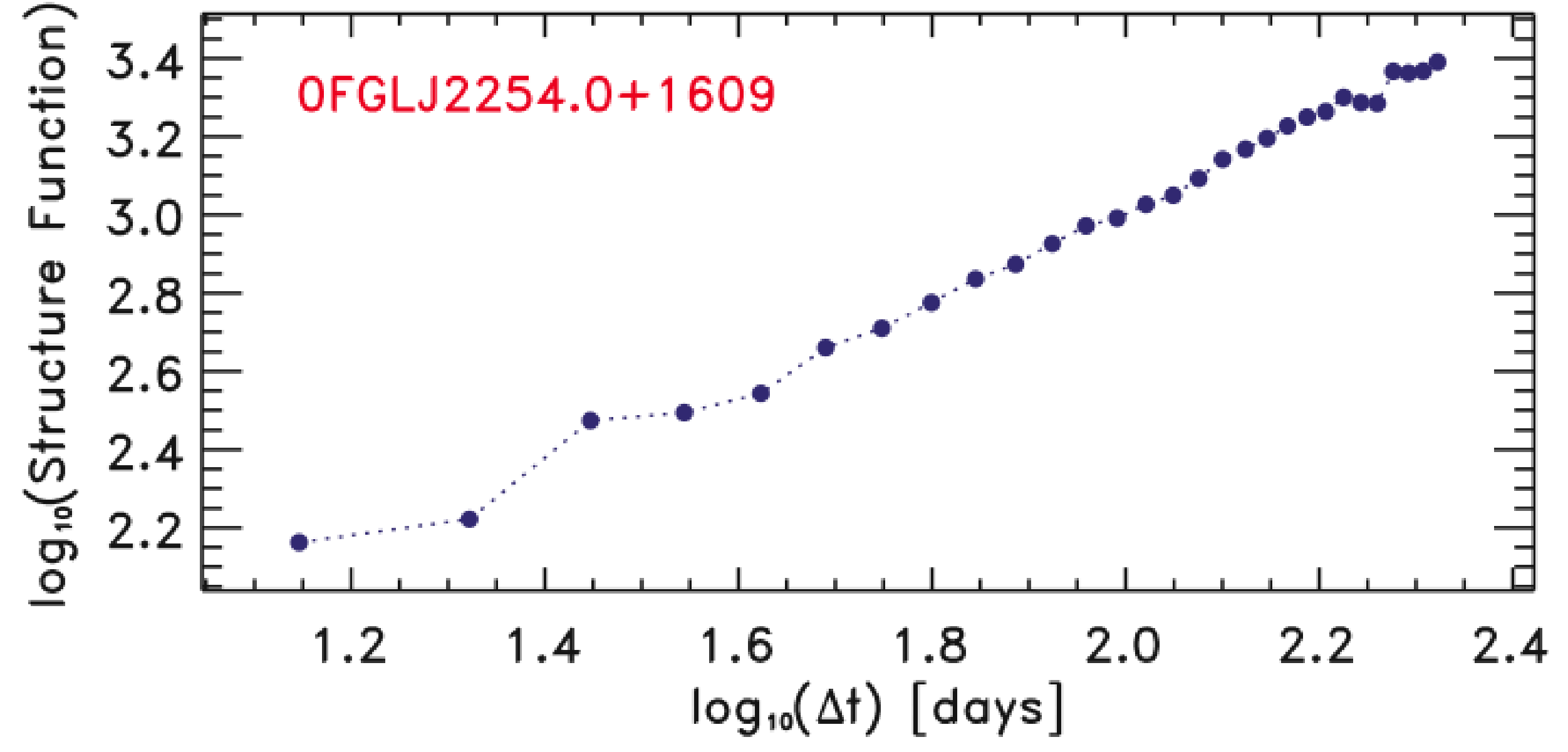}
\caption{\label{label}Structure function for the 1-week-binned LBAS
11 month light curve of 3C454.3 (from [7]).}
\end{minipage}
\end{figure}

\subsection{Structure function analysis}

The structure function is related to both the ACF and PDS. 
The slope of the structure function is related to $\alpha $, the power law
index of the PDS as, $S \propto (\Delta t)^{\alpha -1}$.
Using 1-week-binned light curves, the structure function was computed
for 56 LBAS sources and the corresponding $\alpha $ was estimated for
each source. This is shown in Figure 8 for 3C454.3. For the whole
sample the $\alpha $ values showed a wide distribution with a peak 
at about 1.3.


\subsection{Summary of variability properties}

The Fermi LAT data has allowed us to perform a systematic
study of gamma-ray variability of blazars. 
 
The main results are:
\begin{enumerate}
\item 2/3 of the studied sources are variable. (This fraction is 
increasing with time.)
\item The time spent in high states are $<$ 1/4 of the total observing time.
\item The relative variance is larger for FSRQs than for BL Lacs.
\item ACF time scales are $\approx $ 4 to $> $10 weeks.
\item Typical PDS slopes are $\approx $ 1.3 to 1.5 
\item No evidence for persistent characteristic time scale(s).
\item Flare profiles are on average symmetric
\item The fractional variability during outburst was
found to be similar to its longer term mean in the two objects
where this was studied.
\end{enumerate}

\section{Multiwavelength correlation}
The relation between variability at different wavelengths can be
investigated in a number of different ways. Ultimately
one would like to describe and interpret the detailed time evolution of
the full Spectral Energy Distribution (SED). In practice one
is limited by statistics and by a limited coverage in
wavelength and time. One of the most common tools to extract
information is the Cross Correlation Function (CCF), which averages
correlation information from a whole time series. The
CCF and its discrete version, the Discrete Cross Correlation
Function (DCCF), give the correlation between
two light curves for different relative time shifts or lags.
In the most simple case where one light curve is just a delayed
version of the other, the correlation function will show a peak
at the time lag corresponding to the time shift between the two
bands. In reality the relation between the two bands may be more complex,
including e.g. different time responses and time changes in the
correlation properties. 

In correlation analysis one should also keep in mind that a
detrending will affect the variability timescales to which the analysis is
sensitive. Without detrending the correlation is
often dominated by large amplitude variations on longer time scales.
With a detrending, e.g. by fitting and subtracting a low order
polynomial, these variations are removed and 
variability on shorter time scales will dominate the CCF.
An example of strong correlation between optical and
gamma-ray variations on both long and short time scale
is PKS 0537-441 as can be seen directly in the
light curves of Figure 9. The DCCF for the same data is shown 
in Figure 10. By comparison
the correlation amplitude for the detrended light curves is reduced
from 0.9 to 0.6.

A more complex correlation behavior was seen in PKS 1510-089 [10].
Four flares of about one month each were observed in 2008 to 2009 and
three of these had good coverage in both optical and gamma-rays.
The total DCCF (Figure 11) has a number of correlation peaks.
These are mostly caused by random correlations. Two correlation
features however, appear to be persistent for all the flares.
One is a correlation on short time scales, within 1 - 2 days
of lag zero, The second is a strong peak at a lag of about
2 weeks (gamma-rays leading optical). A detailed comparison
of the light curves show that the envelope (start and end) of
each flare is about the same in the two bands but the
shape of the flare is different. The ratio of gamma-ray
to optical flux is higher in the beginning of the flare
than towards the end. On top of this the flares exhibit
strong variability on shorter time scales which is partly
correlated and responsible for the correlation peak near
zero lag.

\begin{figure}[h]
\begin{minipage}{17pc}
\includegraphics[width=17pc]{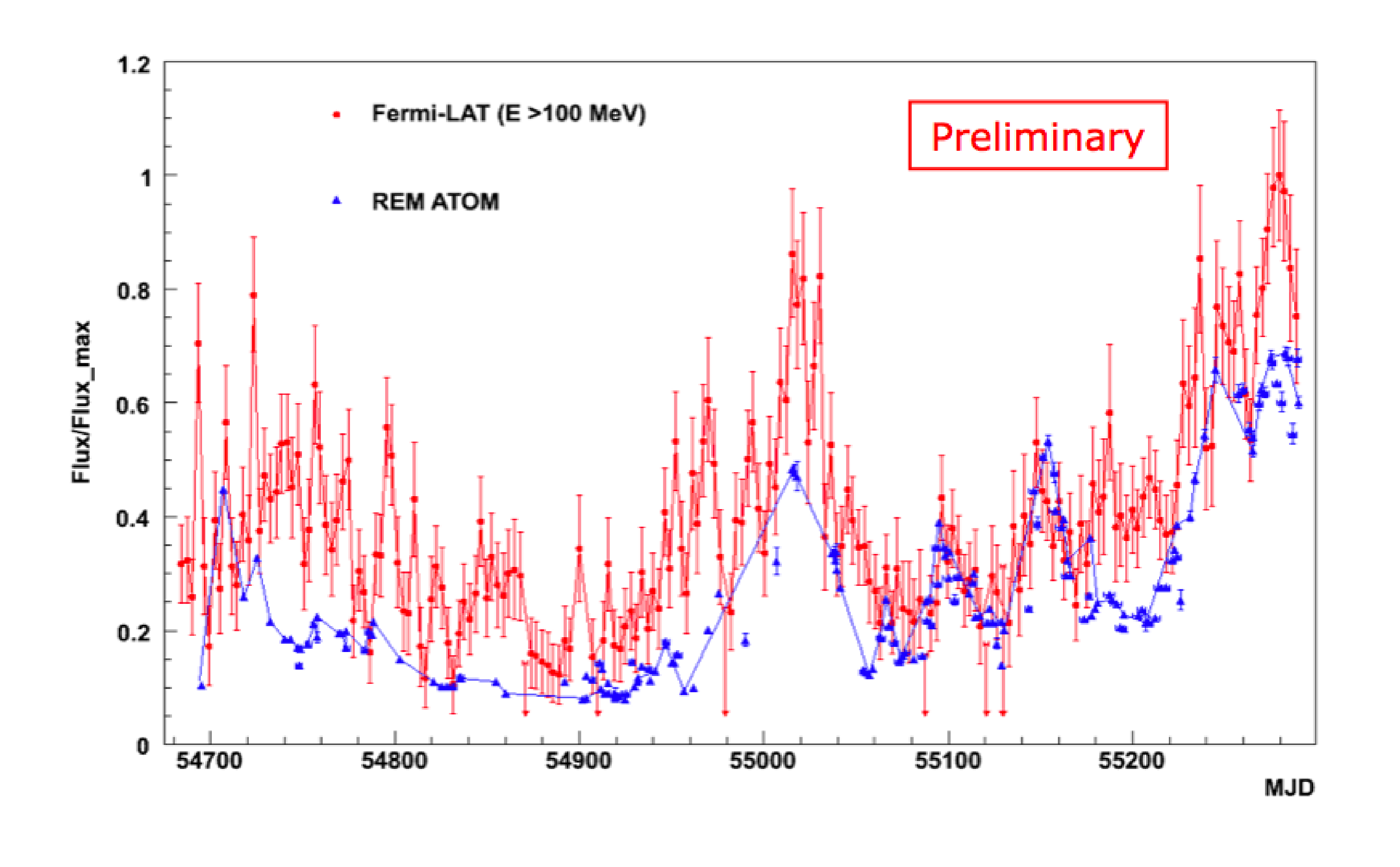}
\caption{\label{label} Comparison between $\gamma$-ray
and R-band light curves for PKS 0537-441. The LAT (red dots) and R (blue
triangles; ATOM and REM) light curves are normalized for comparison
(in preparation).
}
\end{minipage}\hspace{3pc}%
\begin{minipage}{17pc}
\includegraphics[width=16pc]{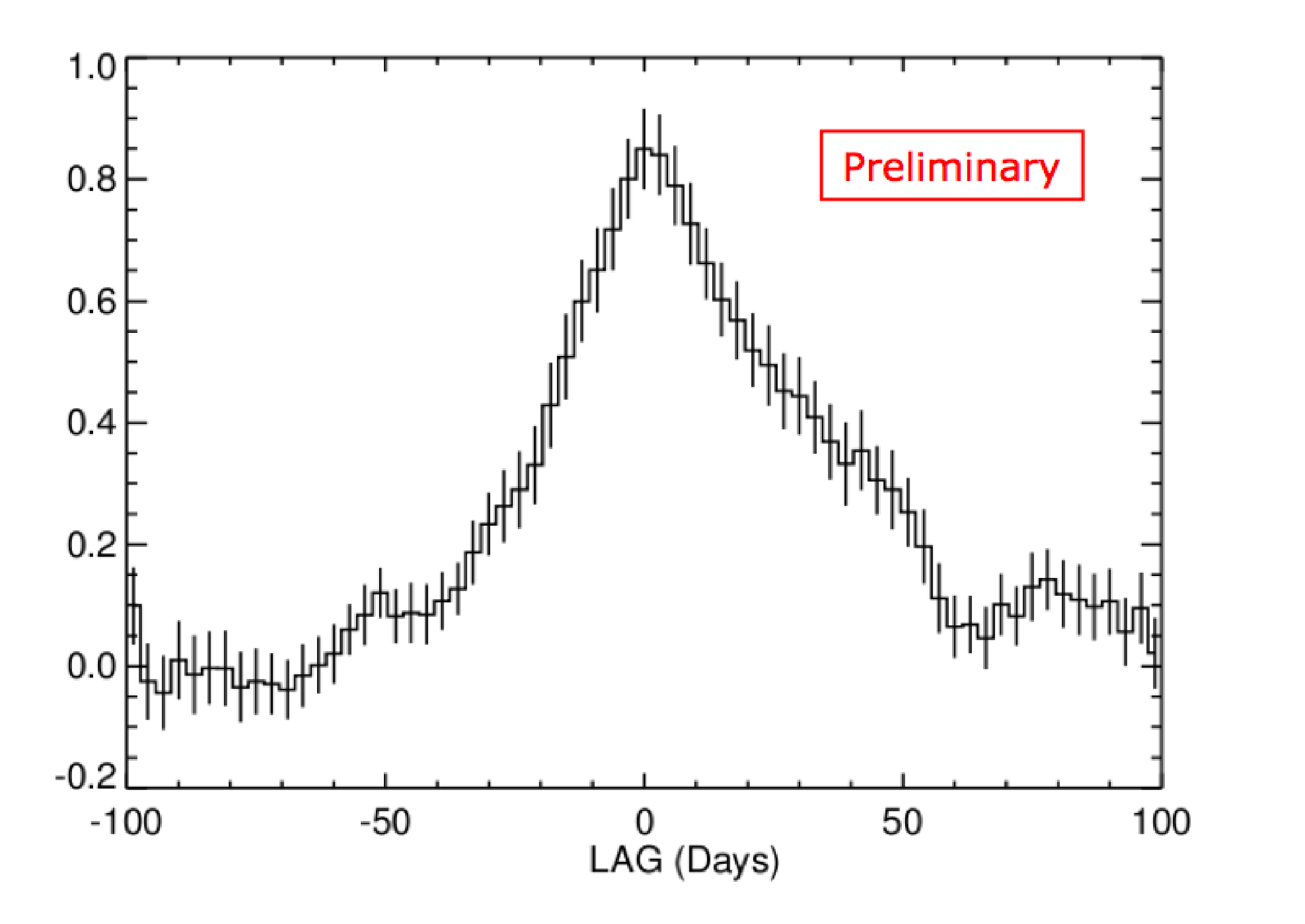}
\caption{\label{label}DCCF between the Gamma-ray and R band light curves 
in Figure 9.
\vspace{0.9cm}
}
\end{minipage} 
\end{figure}

\begin{figure}
\begin{center}
\includegraphics[scale=0.45]{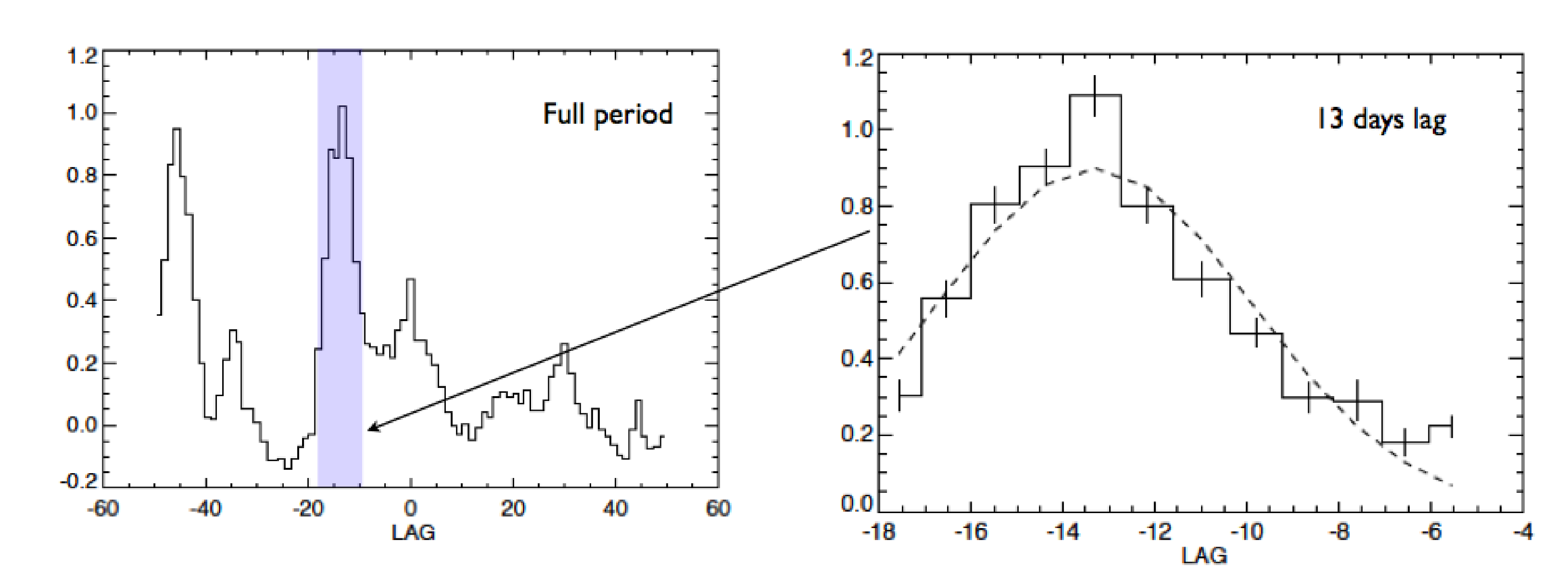}
\end{center}
\caption{\label{label}DCCF between gamma-rays and the R band for 
a period with three major flares in PKS1510-089 (from [10]). 
In a short light curve with just a few flares, there will
also occur correlations at lags corresponding to the time
between the flares. This is the cause of the peak at lag -50
days in this case.}
\end{figure}

\section{Status and future prospects}

The effort to characterize the time variability of 
gamma-ray emission from blazars is presently limited mainly by 
the length of available observations. Variability properties on
short time scales can appear to be quite different from 
the longer term mean due to the stochastic nature of the variability.
Likewise, the multiwavelength properties of flares are not
constant. This is true not just when comparing different objects 
but also for different flares in the same object.

To reveal the nature of the stochastic process or processes
driving the observed variability it is essential to
collect long sequences of observations. Multiwavelength observations 
are just as important in order to identify physical processes
and their connection to each other and to the central engine
at the heart of the blazar and its jet.
Fermi is playing a key role, not just by the unique data it is providing
but also as a catalyst and reference for all other observations
of blazars.
The present decade has all the prerequisites to become a golden
age of blazar research.

\section{Acknowledgments}

The $Fermi$ LAT Collaboration acknowledges support from a number of agencies and institutes for both development and the operation of the LAT as well as scientific data analysis. These include NASA and DOE in the United States, CEA/Irfu and IN2P3/CNRS in France, ASI and INFN in Italy, MEXT, KEK, and JAXA in Japan, and the K.~A.~Wallenberg Foundation, the Swedish Research Council and the National Space Board in Sweden. Additional support from INAF in Italy and CNES in France for science analysis during the operations phase is also gratefully acknowledged.
Larsson is grateful to the Swedish National Space Board for funding
his work in the Fermi project.

\smallskip

\end{document}